\documentclass[preprint2]{aastex}
\usepackage{emulateapj5}
\usepackage{onecolfloat5}
\shorttitle{Deconvolving Pulse Broadening from Radio Pulses}
\slugcomment{Accepted by the Astrophysical Journal, 22 Oct 2002}
\input{clean.defs}
\def\DM{{\rm DM}}     

\def\SM{{\rm SM}}

\newcommand{\taud}{{\tau_d}}
\newcommand{\cnsq}{{C_n^2}}
\newcommand{\Wtau}{{W_{\tau}}}

\newcommand{\be}{\begin{eqnarray}}
\newcommand{\ee}{\end{eqnarray}}

\begin{document}
\twocolumn[
\title{\bf A CLEAN-based Method for Deconvolving Interstellar Pulse
Broadening from Radio Pulses}
\author{N. D. R. Bhat}
\affil{NAIC, Arecibo Observatory, HC 3 Box 53995, PR 00612}
\email{rbhat@naic.edu}
\author{J. M. Cordes \& S. Chatterjee}
\affil{Dept. of Astronomy and NAIC, Cornell University, Ithaca, NY 14853}
\email{cordes,shami@astro.cornell.edu}

\begin{abstract}

Multipath propagation in the interstellar medium distorts radio
pulses, an effect predominant for distant pulsars observed at low
frequencies.  Typically, broadened pulses are analyzed to determine
the amount of propagation-induced pulse broadening, but with little
interest in determining the undistorted pulse shapes.  In this paper
we develop and apply a method that recovers both the intrinsic pulse
shape and the pulse broadening function that describes the scattering
of an impulse.  
The method resembles the CLEAN algorithm used in synthesis imaging
applications, although we search for the best pulse broadening
function, and perform a true deconvolution to recover intrinsic pulse
structre.
As figures of merit to optimize the deconvolution, we use the
positivity and symmetry of the deconvolved result along with the mean
square residual and the number of points below a given threshold.  Our
method makes no prior assumptions about the intrinsic pulse shape and
can be used for a range of scattering functions for the interstellar
medium.  It can therefore be applied to a wider variety of measured
pulse shapes and degrees of scattering than the previous approaches.
We apply the technique to both simulated data and data from Arecibo
observations.

\end{abstract}
]


\section{Introduction}
\label{s:intro}

Multipath propagation in the interstellar medium (ISM) at radio
wavelengths is caused by diffraction from irregularities in electron
density on scales smaller than the Fresnel scale, $\sqrt{\lambda D}
\sim 10^{11}$ cm (where $\lambda$ is the wavelength of observation and
$D$ is the extent of the scattering medium).  One consequence is that
pulses from radio pulsars are distorted by the differences in arrival
time of different portions of the scattered wavefronts.  The resultant
pulse broadening \citep[e.g.][]{will72} is usually quantified in terms
of the pulse-broadening time, $\taud$, which is related to the
scattering measure SM, where SM is the integrated value of $\cnsq$,
the coefficient of the wavenumber spectrum of electron density
irregularities.  Pulse broadening is important information for
quantifying the Galactic structure of electron density fluctuations.
But it is also a hindrance in studying the intrinsic properties of
pulsars that are manifested in the undistorted pulse shape.

For a Kolmogorov wavenumber spectrum with small inner scale
\citep[e.g.][]{rickett1990,CL1991,ARS1995},
the pulse broadening time scales as 
\be
\taud &\approx& 1.10\, {\rm ms}\,\,   \Wtau\SM^{6/5}\nu^{-22/5}D \label{eq:taud},
\ee
where $\nu$ is the frequency of observation (GHz),
$D$ is the pulsar distance (kpc), SM has units kpc m$^{-20/3}$,
 and $\Wtau$ is a geometric 
factor that accounts for the distribution
of $\cnsq$ along the line of sight \citep[e.g.][]{CR1998}.
Pulsars with larger dispersion measures ($\DM$) tend to have larger 
scattering measures.
The effect is thus predominant in observations of 
distant pulsars and at lower frequencies, and it limits the utility of
such scattered pulse profiles for the study of pulsar magnetospheres. 

The observed pulse $y(t)$ can be modeled as a convolution of the
intrinsic pulse $x(t)$ with the pulse-broadening function (PBF) 
$g(t)$ and the net instrumental resolution function, $r(t)$:
\be
y(t) = x(t) \otimes g(t) \otimes r(t);
\label{eq:basic}
\ee
the PBF $g(t)$ is the response of the interstellar medium (ISM)
to a delta function.
The exact form of the PBF depends on the spatial distribution of scattering
material along the line of sight and on its wavenumber spectrum
\citep[e.g.][]{will72,will73,CR1998,LR1999,CL2001,boldyrev-gwinn2002}.

In general, reconstruction of the original pulse shape from the scattered 
data is complicated by the fact that the 
PBF $g(t)$ is unknown. However, classes of possible
PBFs corresponding to different distributions of scattering media
are known.

Independent determination of the PBF requires 
knowledge of the wavenumber spectrum and distribution of scattering 
material along the line of sight.  Alternatively, {\it a priori} 
knowledge of the intrinsic pulse shape could allow determination of 
the PBF.  In general, however, both $x(t)$ and $g(t)$ are unknown; 
hence the utility of scattered pulse profiles is limited to measuring 
$\taud$ through a template fitting method, which assumes a form for 
the intrinsic pulse shape and a model for the ISM, rather than 
determining these quantities from the data.  The assumed shapes can 
be incorrect, leading to estimates for $\taud$ which are only
approximate, and have poorly quantified uncertainties.

In this paper, we describe a method for reconstruction of the
intrinsic pulse shape as well as determination of the PBF. The basic
idea is derived from the CLEAN algorithm extensively used in
image reconstruction of synthesis imaging data \citep{hogbom74,schwarz78}.  
Our one-dimensional application is similar to the usage of
CLEAN in spectral analysis \citep[]{Roberts87}. 
In \S\ref{s:method}, we outline the method and its practical
realization, and compare it to other methods that have been used for
this application.  This CLEAN-based method is demonstrated using 
simulated data as well as some recent data from Arecibo observations, 
and the results are discussed in \S\ref{s:results}. Finally, we
present our conclusions in \S\ref{s:conclusions}.

\section{Method and Practical Realization}
\label{s:method}

\noindent
Before outlining the CLEAN-based approach, we briefly describe other
methods that have been used to 
estimate the pulse-broadening time from scattered pulse profiles.    
In general, a model is fitted 
to the observed pulse profile, 
$\yo(t)$, where the model, $\ym(t)$, is the convolution of a 
template profile, representing an estimate for the intrinsic pulse shape, 
with an impulse response, $g(t)$, that describes pulse broadening
from the ISM  for the particular line of sight.  

\noindent
\subsection{Conventional Frequency-Extrapolation Approach}
\label{s:conv}
Typically,
the intrinsic pulse shape at the frequency of interest
 is extrapolated from high-frequency observations
where the pulse broadening is assumed to be negligible.  In addition,
the PBF shape is also assumed known, often as a one-sided
exponential function, 
\be
g(t) = \taud^{-1}\exp(-t/\taud)U(t), 
\label{eq:thin}
\ee
where
$U(t)$ is the unit step function.
As discussed below, this form applies to a thin scattering
screen for which the density irregularities follow a square-law
structure function.
The pulse broadening time $\taud$ is estimated by minimizing
$\chi^2$ (as standardly defined) as a function of $\taud$, 
the sole parameter in this problem. 
Results from the method thus rely on use of an empirical pulse shape
derived from radio frequencies quite different from that at which the
scattering is prominent.  However, it is well known that profile
evolution with frequency can be quite complex for some pulsars
\citep[e.g.][]{hankins-rickett1986}, and extrapolation to the
frequency of interest will be in error.  It is also well known that
the PBF is a one-sided exponential only for a medium described as a
thin screen with a square-law structure function
\citep[e.g.][]{CR1998}.  While thin screens may exist along some lines
of sight, a more reasonable default model is one where the medium is
Kolmogorov (leading to a non-square law structure function) and fills
a significant fraction of the line of sight.  Despite such
shortcomings, the method has been used to obtain estimates of $\taud$
for a large number of pulsars \citep[e.g.][]{lohmer2001}.

Variations on this approach include use of non-exponential
pulse broadening functions and detailed estimation of the
frequency dependence of the intrinsic pulse shape, for example
by modeling individual components of the pulse shape and
scaling each with frequency.   The essence of the method
is the same, however, in that measurements at other frequencies
are used to estimate the intrinsic pulse shape at the frequency
of interest.

\noindent
\subsection{Fourier Inversion}  
\label{s:four}
From Eq.~\ref{eq:basic}, the Fourier transform of $y(t)$ is related 
to that of $x(t)$ as $Y(f)=X(f)G(f)R(f)$, where $X,G$ and $R$ are the
respective Fourier transforms of the factors on the right-hand side of
Eq.~\ref{eq:basic}.
Adopting a particular form for $g(t)$ and knowing $r(t)$, one
can, in principle, perform the deconvolution by calculating
 $X(f) = Y(f)/G(f)R(f)$  and inverse-transforming to 
obtain the intrinsic pulse shape.
This method was used by \citet{weisbergetal1990} and \citet{kuz93}
with a one-sided exponential PBF.  
Kuz'min \& Izvekova pointed out that the results
depend on the assumed form of the PBF but did not present any
results using alternative forms.    

\noindent
\subsection{A CLEAN-based approach}
\label{s:clean}
In this paper, we propose an alternative method to realize both objectives 
of recovering the intrinsic pulse shape and determining the 
shape and characteristic time scale of the broadening function.
The method is based on the CLEAN algorithm, an iterative method 
for deconvolving the instrumental point source response function from 
an interferometric image \citep{hogbom74,schwarz78}. In the present case, 
the problem is much simplified as the data are one-dimensional.
The method involves inverting the convolution in Eq.~\ref{eq:basic}. 
By analogy with imaging applications, the
measured, scattered pulse is equivalent to the ``dirty
map''; and the combination $g(t)\otimes r(t)$ 
(the PBF convolved with the instrumental response) is 
equivalent to the ``dirty beam''. 
In synthesis imaging applications, CLEANing involves the iterative
subtraction of scaled copies of the dirty beam from the dirty map at
locations corresponding to real features until the residual map is
indistinguishable from noise, followed by restoration of the
subtracted CLEAN components. Here, we adapt the method to 
(one-dimensional) scattered pulse profiles. 


While this method is inspired by CLEAN as used in synthesis imaging,
we emphasize that there are some important differences between the two
algorithms.  Specifically, in imaging applications, the dirty beam is
assumed to be perfectly known, and sets an upper limit on the
achievable resolution.  In the algorithm outlined here, we are
performing a true deconvolution, and the resolution of the deconvolved
pulse is determined by the response function of the instrument, rather
than the width of the PBF.  Additionally, the PBF is not known {\em
a priori}, and we perform a search in both parameter space and
function space for the best pulse-broadening function, as outlined below.

We consider the measured pulse to consist of delta functions, each
convolved with the PBF, $g(t)$, and the resolution function, $r(t)$.
In reality, radiometer noise adds to the measured quantity and the
intrinsic pulse shape is itself perturbed by pulse-to-pulse
fluctuations intrinsic to the pulsar.  The CLEAN process identifies
the amplitude and location of a CLEAN component (CC) by finding a suitable
maximum in the residual profile in each iteration.  Initially, the
residual profile is the measured profile, $y(t)$.  The amplitude of the 
CC is the identified maximum multiplied by a {\it
loop gain}, $\gl$.
The CC \yc(t) is convolved with $g(t)$ and $r(t)$ 
and subtracted from the measured pulse profile to
yield the  residual pulse profile, $\yd(t)$: 
\begin{equation}
\yd(\ti) = y(\ti) - \left[ \yc(t) \otimes \left(g(t) \otimes
r(t)\right) \right]_{t=t_i}, \quad i=1,...,N,
\end{equation}
where $\yc(\ti) = \gl \{{\rm max} [y(t)]\} \delta (t-\tnot)$, and
$N$ is the number of pulse phase bins. $\yd(t)$ for a given
iteration becomes $y(t)$ for the next iteration, and the process is
repeated until the residual subtracted pulse, $\yd(t)$, is
comparable to the off-pulse rms noise level\footnote{The termination
criterion is not well-defined for the CLEAN technique.}.  Upon
termination of the iteration, $\nc$ CLEAN components are identified:
$C_j,t_j$, $j=1,\ldots,n_c$ denote the amplitudes  and times 
of these CLEAN components (i.e., the collection of delta functions
at the end of the CLEAN process).

\begin{figure*}[htf]
\epsscale{1.75}
\plottwo{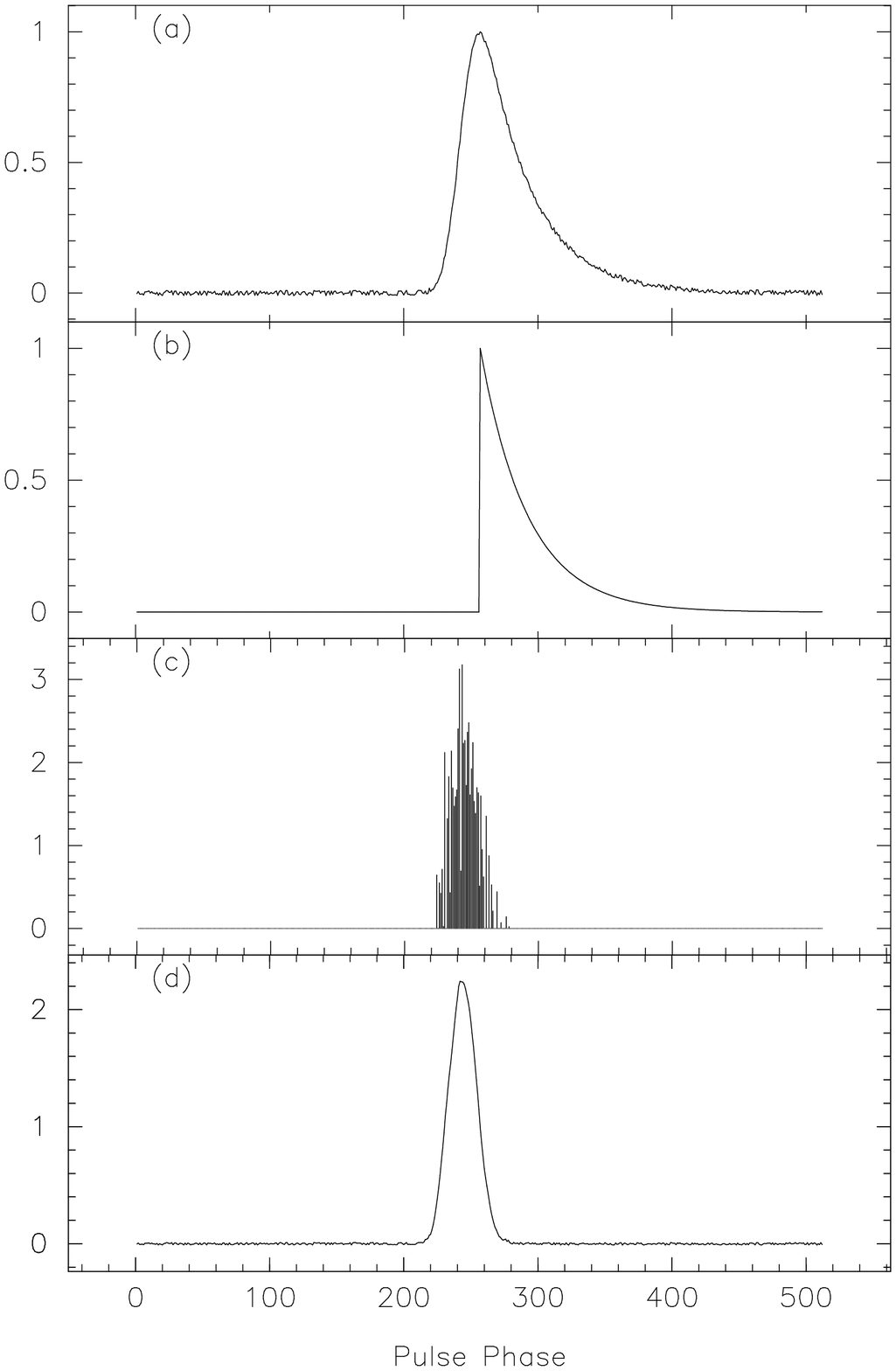}{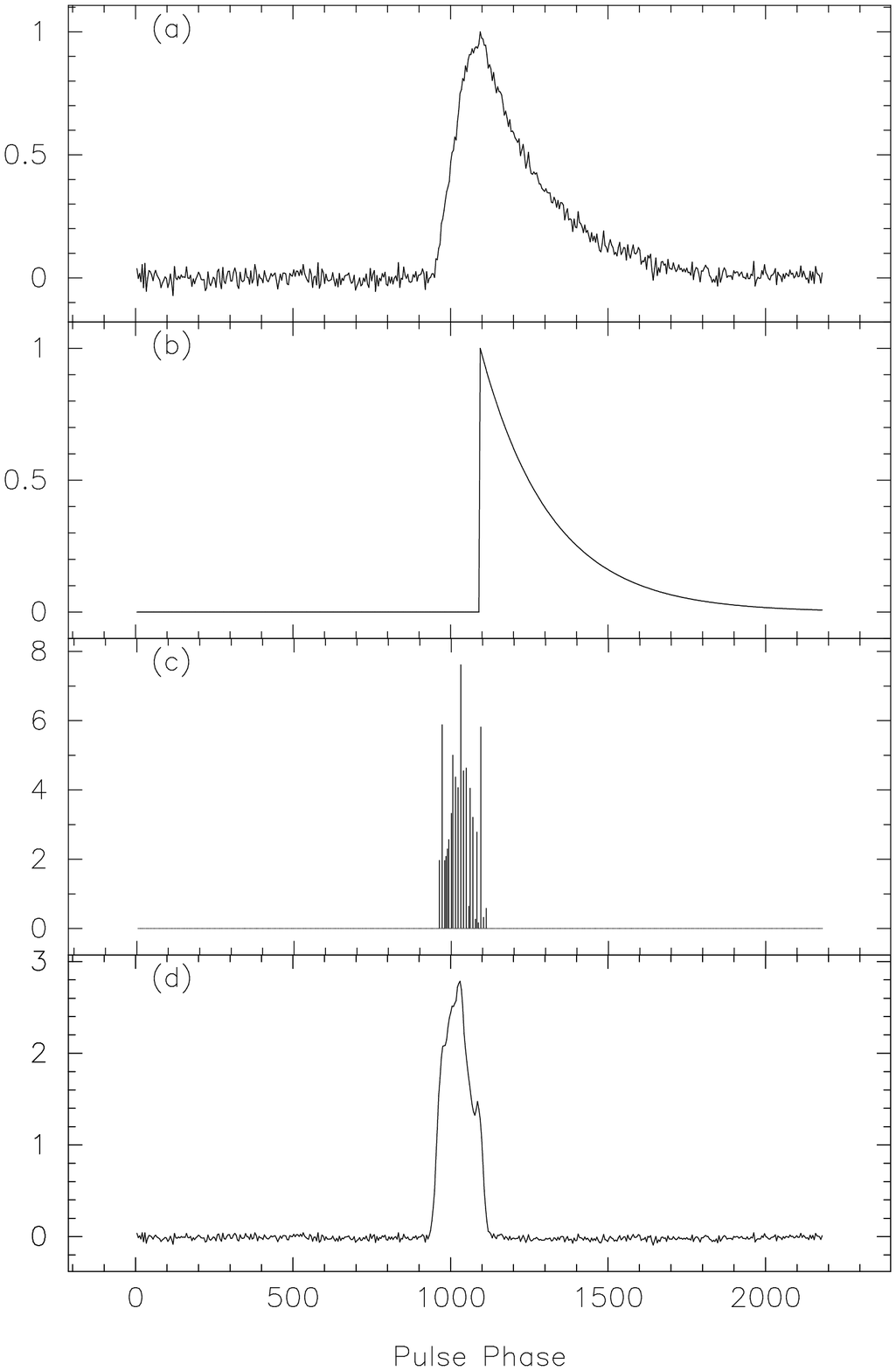}
\caption{Sample plots that illustrate the CLEAN process for
deconvolving the interstellar pulse broadening.
Left-hand side: simulated data.  Right-hand side: data for
PSR J1852+0031 at 1475 MHz.
(a) scattered pulse shapes
(b) model PBFs
(c) CLEAN components
(d) final restored (CLEANed) pulse shapes.}
\label{fig:demo}
\end{figure*}

The restored pulse shape is built from the ensemble of CLEAN
components, $C_j$, convolved with a `restoring function'
$\rho(t)$, details of which we will discuss later in this section:
\begin{equation}
\yr(t) = \sum _{j=1}^{\nc} C_j\delta(t-t_j) \otimes \rho(t).
\end{equation}
At the end of the CLEAN process the residual noise, $\yd(t)$, is added
to the convolved CCs to obtain the final restored pulse profile.

Examples of the CLEAN process are shown in Figure~\ref{fig:demo} for
simulated data (left-hand panels) and real data (right-hand panels).
The figure shows the measured pulse shapes, the chosen PBF (a
one-sided exponential as in Eq.~\ref{eq:thin}), the CCs, and the
deconvolved profile.  Compared to the traditional least-squares or
frequency-extrapolation methods (\S\ref{s:conv}), the approach to the
pulse-broadening problem outlined here is much easier to apply because
it is non-parameteric. Further, it does not make any assumption about
the intrinsic pulse shape or its evolution with frequency, and thus
can potentially yield useful information about the intrinsic pulse
shape.


Below we elaborate on some aspects concerning the practical
realization of the method.

\noindent
\subsubsection{Loop Gain}
\label{s:loop}
Small values of the loop gain ($\gamma$) are employed in imaging
applications. The performance of the method is expected to change only
marginally for sufficiently small values (say, $\gl \ll 0.1$).  Larger
values tend to cause over-subtraction and consequent lack of
convergence.
We adopted \gl=0.05 as a reasonable choice for our application.  Thus,
in each iteration, a delta function with an amplitude of 5\% of the
profile peak is convolved with the chosen PBF and the instrumental
response function $r(t)$, and then subtracted from the peak of the
residual pulse profile.

\noindent
\subsubsection{ Pulse Broadening Functions} 

Unlike radio synthesis imaging applications, where the dirty beam is 
often known exactly as the response of the instrument to a point source, 
the exact functional form for the PBF of the ISM is not known.
It depends on both the distribution  and the wavenumber spectrum of 
scattering material along the line-of-sight to the pulsar, neither of
which is known in detail. 
The PBF is particularly sensitive to the spatial distribution
of $\cnsq$. 
For instance, in the simplest case of a thin slab 
of scattering material located in 
between the pulsar and the Earth, the PBF is characterized by a sharp rise 
and exponential fall-off (eq.~\ref{eq:thin}), whereas the PBFs of more 
realistic models (such as a thick slab or a uniform medium) are 
characterized by rounded edges and slower fall-offs, with different rise 
and peak times.
For media with a square-law structure function 
\citep{CR1998, LR1999} 
the PBF for a thin screen is given by Eq.~\ref{eq:thin} and for
thick and uniform media the PBFs are given by
\citet[][]{will72,will73}:
\be
g(t) &=& \left( \pi \taud / 4 t^3 \right)^{1/2} \exp(-\pi^2\taud/16t) \\
\label{eq:thick}
g(t) &=& \left( \pi^5 \taud^3 / 8 t^5 \right)^{1/2} \exp (- \pi^2 \taud / 4t)
\label{eq:uniform}
\ee
where the time constant $\taud$ is the pulse-broadening time. 

For other media, such as those with a Kolmogorov wavenumber spectrum 
having different inner scales 
\citep{LR1999} or media with non-Gaussian density fluctuations
\citep{boldyrev-gwinn2002}
(but still having uniform statistics transverse to the line of sight),
the PBFs are qualitatively
similar to those for media with square-law structure functions in that
thin screens show sharp rise times while extended media have slower
rise times.

Nonuniformities of the scattering transverse to the line of sight,
particularly any truncation of the scattering screen, produce
PBFs truncated for times $t > \tmax$, where $\tmax$ is the arrival
time corresponding to the edge of the screen 
\citep{CL2001}. 
As an example, a circular disk centered on the line of sight
to a source will produce a PBF of the form
\begin{equation}
g(t) = { \exp (-t / \taud) \over \taud (1 - \exp (-\zeta)) } U (\tmax - t),
\end{equation}
where $\tmax = \zeta \taud$, $\zeta = \thetamaxsq / 2 \sigthetasq $, where
$\thetamax$ and $\sigtheta$ are the maximum and RMS scattering angles, 
respectively. 
\citet{CL2001} also discuss PBFs for several other configurations.
For example, in the case of scattering from a long, narrow filament,
$\thetamax$ will be much larger than $\sigtheta$, and consequently the
scattered image becomes elongated, and the PBF tends toward the form
$\propto t^{-1/2} \exp(-t)$.  For a
large screen with elongated irregularities, the PBFs will be strongly
frequency dependent, and will be characterized by multiple peaks that
align at different frequencies.

In applying our algorithm, one can perform a grid search over
various forms of the PBF as well as the relevant parameters and
thus determine 
(a) the appropriate model of the PBF that best describe 
observations, and (b) the best fit $\taud$ that yields meaningful 
intrinsic pulse shapes. The figures of merit for 
determining the best fit case is discussed later in this section. \\

\noindent
\subsubsection{ Instrumental Response}

The response function due to instrument can be described as 
the convolution of the functions that characterize various 
instrumental effects relevant in determining the effective 
pulse phase resolution, including
(i) dispersion smearing ($\rdm$), 
(ii) profile binning ($\rpb$), (iii) the backend time 
averaging ($\rav$), and (iv) additional post-detection time averaging
($\rpd$). The response function of the instrument, or the 
effective resolution function of the data, is therefore 
given by
\begin{equation}
r(t) = r_{dm}(t) \otimes r_{pb}(t) \otimes r_{av}(t) \otimes r_{pd}(t).
\label{eq:rstfn}
\end{equation}
The contributing factors in the above equation, in particular those 
which describe effects (i) to (iii), can be treated as 
approximately rectangular.
Hence the response function will be approximately ``trapezoidal'' for 
cases where one or two factor(s) dominate the effective time resolution 
of the data. It will be more ``Gaussian-like'' due to the convolution 
of similar-sized rectangle functions when there are multiple 
contributions that determine the time resolution. We form the
resolution function according to eq.~\ref{eq:rstfn},
where the convolution is performed with a time resolution that is much 
larger than the narrowest factor; the resultant function is then 
re-sampled to the resolution of the actual data. 

\noindent
\subsubsection{Restoring function} 
\label{s:restor}

The CLEANed pulse shape is obtained by convolving CCs
with a restoring function, $\rho(t)$, equivalent to the
 CLEAN beam in imaging applications. 
The choice of a restoring beam is driven by several ideas (see 
\citet{cornwelletal1999} for a detailed discussion). Following 
the analogy of imaging applications, where the CLEAN beam is 
approximated as a Gaussian fitted to the central part of the dirty 
beam, we adopt an equivalent Gaussian, $\rg(t)$,
whose width reflects the effective resolution 
of the actual data (given by the width of $r(t)$ as discussed in 
the previous section). The choice of restoring function,
$\rho(t) = \rg(t)$,  
de-emphasizes the higher frequency components that are sometimes 
spuriously generated by the CLEAN algorithm when directly using the
response function $r(t)$ as the restoring function.
The restoring function is normalized to unit area in order to 
ensure conservation of flux in the inversion process. 

\noindent
\subsubsection{Figures of Merit}
\label{s:figmer}
We now discuss the figures of merit for determining the best fit case, i.e.,
when the observed pulse profile is optimally compensated for interstellar
broadening.   Assume that we know the shape of the PBF but not its 
characteristic scale, $\taud$.   If we choose $\taud$ too large,
we will tend to oversubtract pulsed flux and thus introduce
negative going features in the deconvolved profile.  Conversely,
if too-small a value for $\taud$ is used, the deconvolved profile
will display residual asymmetry due to un-deconvolved scattering.
Therefore two figures of merit that are appropriate are:
(1) positivity and (2) minimum asymmetry of the resultant pulse.
In practice we numerically require the residual noise
(i.e., measured$-$modeled pulse) to show zero mean and the same rms 
everywhere.  We use all these criteria to choose the
correct PBF and characteristic time scale, $\taud$. 

We define a parameter ($f_r$) that measures positivity as
\begin{equation}
f_r =  \frac{ m}{N \sigoffsq } \sum _{i=1} ^{N} [\yd(\ti)]^2
U _{\Delta y} 
\end{equation}
where 
\be
U _{\Delta y} = U \left( -  \Delta y (\ti) - x  \sigoff  \right),
\ee
$\yd(\ti)$ is the residual noise at the end 
of the CLEAN process, and $m$ denotes a `weight' of 
order unity. The unit step function $ U$ turns on 
when the residual $\Delta y$ is significantly below the off-pulse
noise level (i.e. when oversubraction is caused by using a PBF that is
too broad).  
We adopted $x=3/2$ for our analysis so that 
distortions of the pulse more negative than $3\sigma/2$ 
yield a penalty manifested as a larger value of $f_r$.
To minimize profile asymmetry, we minimize the 
skewness parameter of the CLEANed profile, $\yr(t)$, by explicitly
minimizing the skewness ($\Gamma$) of the CCs:
\be
\Gamma = { \langle t^3 \rangle \over \langle t^2 \rangle ^{3/2} },
\ee
where 
\be 
\langle t^n \rangle &=&
		\frac{\displaystyle
		  \sum_{i=1}^{n_c} (t_i - \bar t)^n C_i
		}
		{\displaystyle
		  \sum_{i=1}^{n_c}  C_i
	        }, \\
\overline {t} &=& 
		\frac{\displaystyle
                  \sum_{i=1}^{n_c} t_i C_i
                }
                {\displaystyle
                  \sum_{i=1}^{n_c}  C_i                
                }, 
\ee 
and $C_i,t_i$, $i=1,\ldots,n_c$ denote the amplitudes  and times 
of the CLEAN components, as defined before.

In addition to the symmetry and positivity constraints, we 
also maximize the number of residual
points in the on-pulse window that were consistent with the
noise in an off-pulse window. Specifically we count the number
of points $N_f$ that satisfy 
$\vert \yi - \yoff \vert \le 3\sigoff$,    
where \yoff and \sigoff represent the mean and rms of the off-pulse
region.  Thus the figures of merit we employ include (a) maximum
$N_f$, (b) minimum residual noise, (c) minimal skewness, and
(d) minimum positivity parameter $f_r$.

\begin{figure}[hf]
\epsscale{0.85}
\plotone{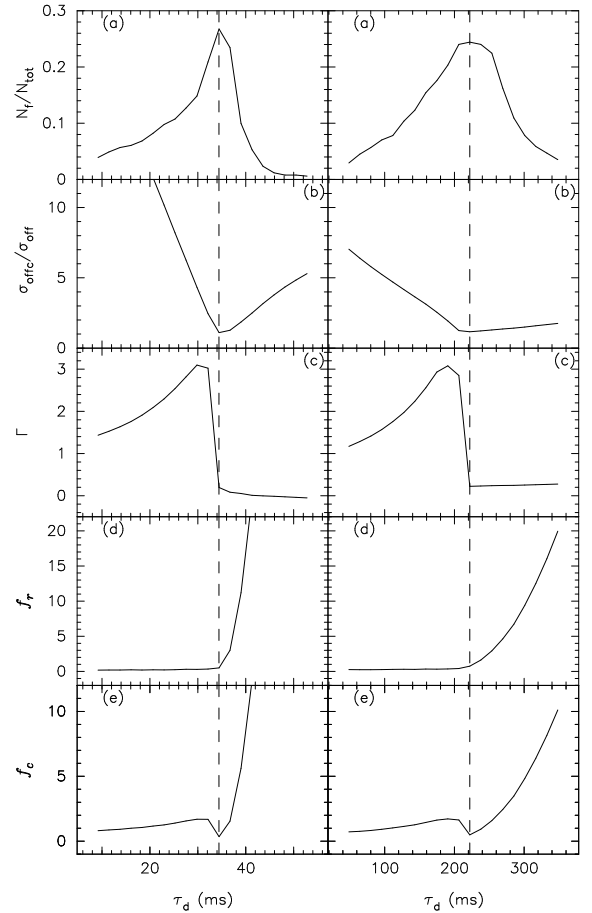}
\caption[]{Plots of the various figures of merit used in the CLEAN
method as a function of the model parameter, $\tau _d$.
Left-hand side: simulated data from Figure~\ref{fig:demo}.
Right-hand side: real data from PSR J1852+0031 at 1475 MHz also
        in Figure~\ref{fig:demo}.
The vertical dashed lines show the true value of $\taud$ for
the simulated case and the best value for the real data.
(a) $N_f$,  the number of data points within 3-$\sigma$ level of the noise,
        as described in the text;
(b) $\sigoffc/\sigoff$, the ratio of RMS residual noise to the off-pulse
RMS noise;
(c) the skewness parameter, $\Gamma$;
(d) the positivity parameter, $f_r$;
and
(e) the combined skewness + positivity parameter,
$f_c = (\gamma + f_r)/2$
}
\label{fig:bestfit}
\end{figure}

Sample plots illustrating the figures of merit
are shown in Figure~\ref{fig:bestfit}  for the simulated and actual
data of Figure~\ref{fig:demo}.
The best fit model will correspond to minimal values of both $f_r$
and $\Gamma$, or equivalently to a minimum of the combined parameter,
$f_c = (\Gamma + f_r)/2$, as well as conforming to a maximum of $N_f$
and a minimum of the residual noise rms (\sigoffc) of the CLEANed pulse. 
Economy in the number of CCs required to fit the observed
pulse profile may also provide a means to discriminate between
different forms of the PBF (and hence, different distributions of
scattering material).

As the deconvolution process removes smearing due to interstellar
scattering, the resultant CLEANed pulse is narrower and the peak 
signal to noise (S/N) is higher, the increase being
determined largely by the ratio of $\taud$ to pulse width.
This is evident from examples of simulated and real data shown in
Figs.~\ref{fig:demo}~and~\ref{fig:sim1}--\ref{fig:examples};
examination of Figure~\ref{fig:demo} suggests that the
sum of CCs matches closely the area of the scattered pulse.

\section{Further Examples}
\label{s:results}
We illustrate the application of the above-described method using 
additional simulated and real data.    In doing so, we show
the sensitivity of the method for finding both the shape of the
pulse broadening function and the appropriate $\taud$.    We also
demonstrate how the derived intrinsic pulse shapes vary with
different PBFs. 

\noindent
\subsection{Simulated Data}
\label{s:sim}
\begin{figure}[hf]
\epsscale{1.0}
\plotone{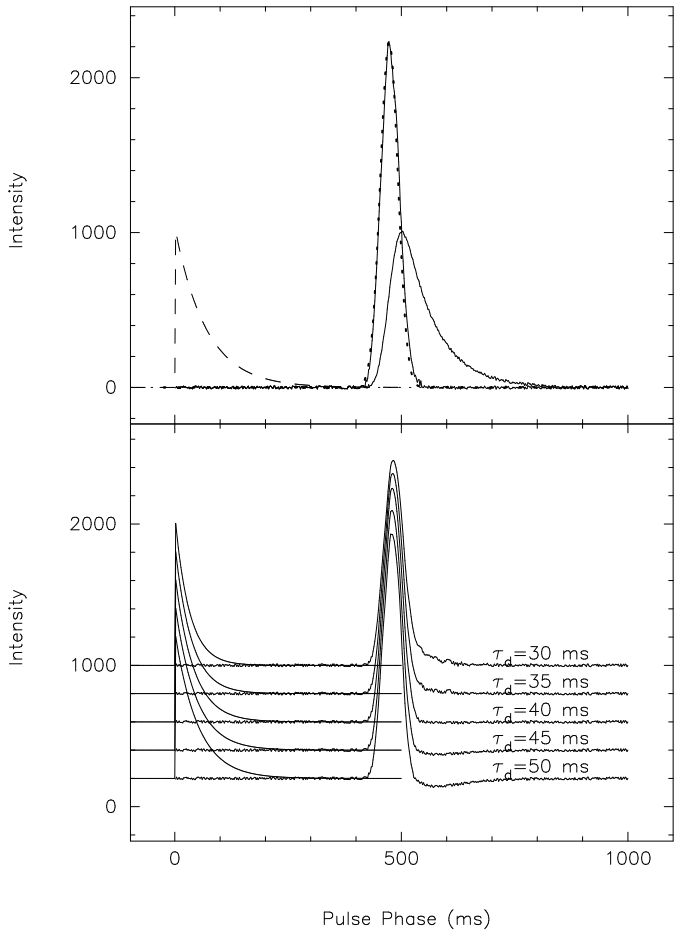}
\caption[]{Example illustrating the CLEAN-based deconvolution
for simulated data and trial PBFs having a range of
pulse-broadening times.
Top frame: the restored and true
pulse shapes are shown as solid and dotted curves, respectively, along with
the true PBF (a one-sided exponential with $\tau$ = 40 ms)
used to generate the scattered pulse (dashed line).
Lower panel: PBFs and resultant deconvolved pulse shapes
for a range of models that have the correct
PBF shape but a range of pulse-broadening times,
$\taud = 30, 35, 40, 45$ and 50 ms, from top to bottom trace.
PBFs that are narrower than the true PBF result in pulse shapes with
residual scattering ($\taud$=30 and 35 ms), while wider PBFs
over-subtract the data, causing negative features near the
trailing edge of the pulse ($\taud$=45 and 50 ms).}
\label{fig:sim1}
\end{figure}

Pulse profiles are generated as  single or multi-component 
Gaussians. Scattered profiles are then constructed 
for a given choice of PBF, and noise 
is added to the scattered profile to account for finite signal-to-noise
ratios. The simulated scattered data are then input to our
CLEAN algorithm to recover the intrinsic pulse shape as well as the 
best fit model PBF, and the results are compared with the original 
data and PBF used to generate the data.
Figure~\ref{fig:sim1} illustrates the case
where the intrinsic pulse is a simple Gaussian and the ISM is
described by a simple thin slab model. 
As evident from the figure,
the intrinsic pulse
obtained by application of the CLEAN algorithm is in very good agreement 
with the original data. A significant deviation from the best fit model 
PBF (a one-sided exponential with a characteristic timescale of 40 ms) 
leads to distortion of the resultant deconvolved pulse profile, 
leading to residual scattering (for PBFs narrower than the optimal
case) or negative going features at trailing edge of the pulse (for
PBFs wider than optimal case). This example thus demonstrates the 
ability of the technique to well characterize the PBF and recover 
the intrinsic pulse.

\begin{figure}[hf]
\epsscale{1.0}
\plotone{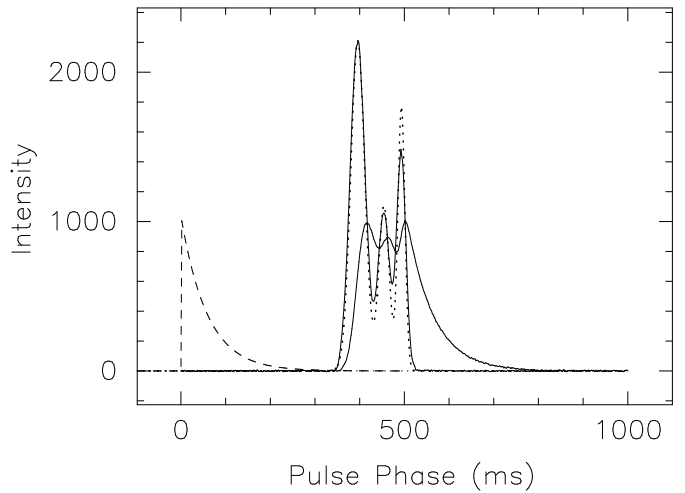}
\caption[]{Simulation example illustrating the CLEAN-based technique for
recovering an intrinsic pulse shape consisting
of three components. The CLEANed pulse (solid line) is overlayed
on the original pulse (dotted line); the best fit PBF model (exponential
with a time constant of 60 ms) is shown on the left (dashed curve).}
\label{fig:sim2}
\end{figure}

Fig~\ref{fig:sim2} shows another example where the intrinsic pulse is
modeled as a sum of Gaussians, to mimic a pulse profile that is 
composed of multiple components. As in the previous example, the
ISM model assumed here is a 
thin slab of scattering material, which can be described by an 
exponential PBF with a
characteristic time scale $\taud = 60$~ms. 
The recovered pulse is in good agreement with the original data.


\noindent
\subsection{Data from Observations} 
\label{s:data}
\begin{figure}[hf]
\epsscale{0.95}
\plotone{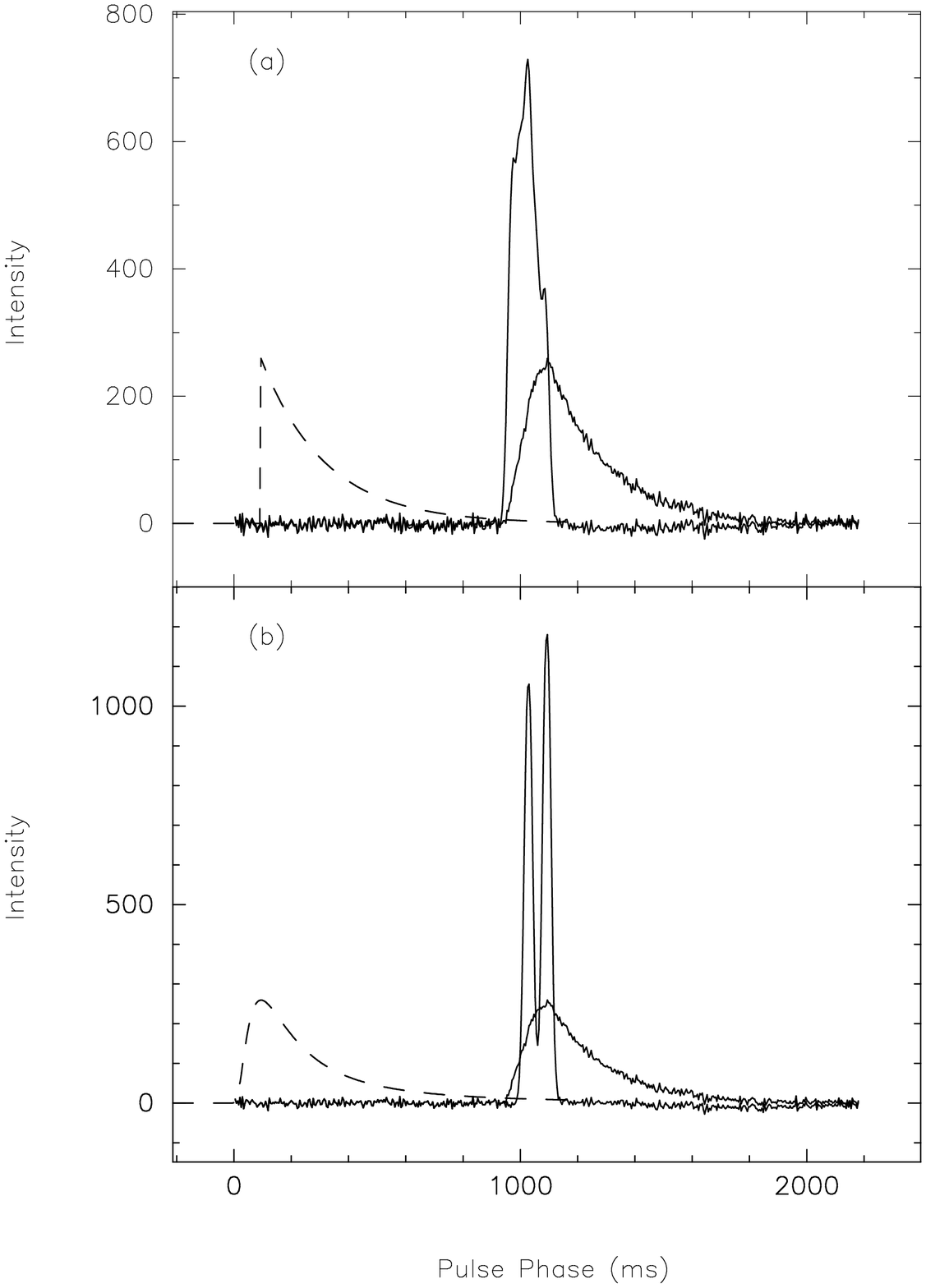}
\caption[]{CLEANed pulse shapes for PSR~J1852+0031 at 1475 MHz
obtained by deconvolving with two different models for the PBF. The
results are strikingly different.  Upper panel: a one-sided
exponential PBF, corresponding to a thin-screen model with a
square-law structure function, yields $\taud$=225$\pm$14 ms and an
approximately triple pulse with merged components.  Lower panel: a
more rounded PBF, corresponding to a uniform medium with a square-law
structure function, yields $\taud$=121$\pm$6 ms and a distinct double
pulse shape.  The differences result from the fact that in the top
panel, the sharp PBF requires broad structure in the deconvolved pulse
in order to match the measured pulse.  For the bottom panel, the PBF
is more rounded and the deconvolved pulse can have sharper structure
and still be consistent with the measured pulse.  }
\label{fig:J1852}
\end{figure}

\begin{figure*}[hbf]
\epsscale{1.85}
\plotone{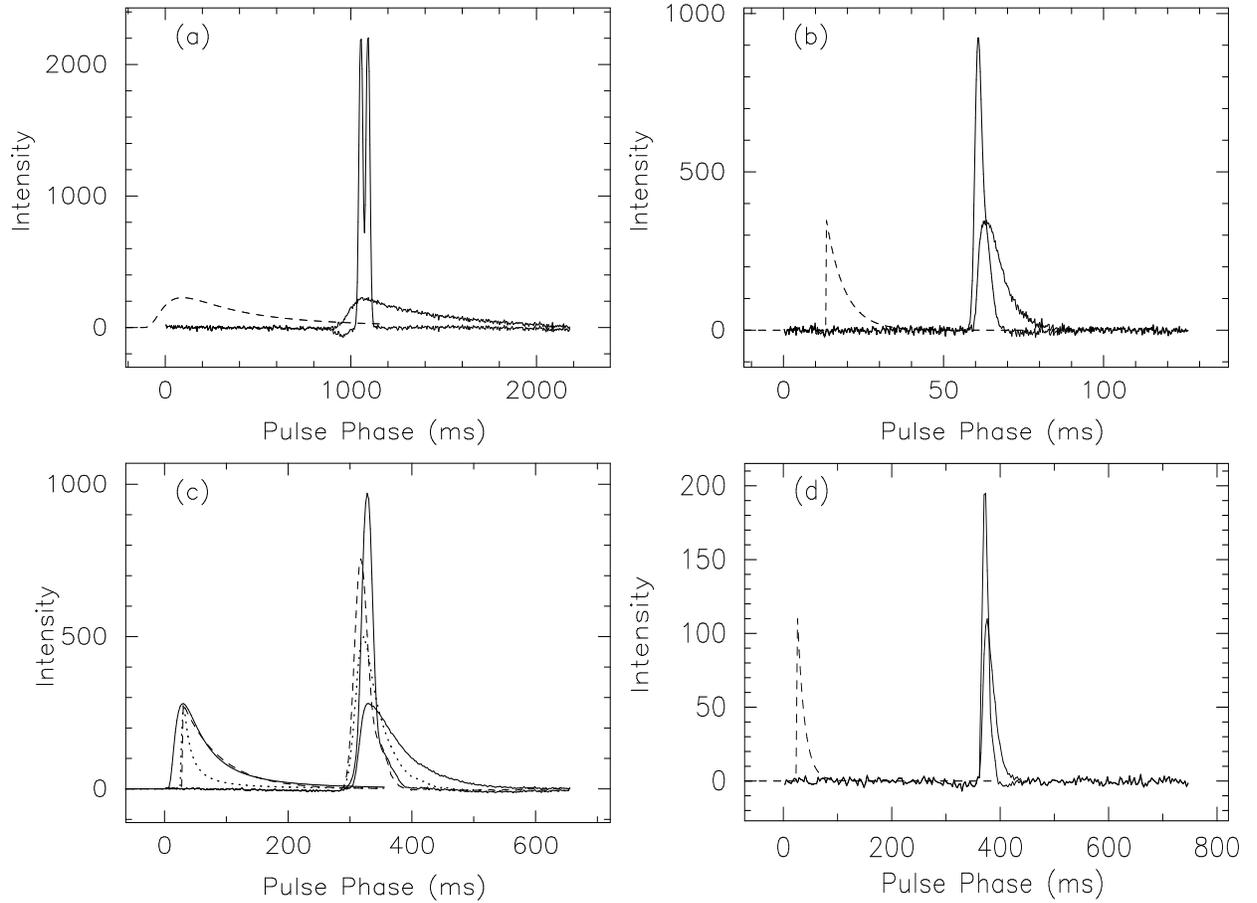}
\caption[]{The measured  and CLEANed  pulse profiles for four pulsars:
(a) J1852+0031 at 1175 MHz,
(b) J1853+0546 at 1475 MHz,
(c) PSR J1901+0331 at 430 MHz;
The measured pulse can be accounted for by
either a thin slab model with a PBF of characteristic time,
$\taud$=60$\pm$3 ms (dashed curves), or by a uniform medium
with a PBF with $\taud$=31$\pm$2 ms (solid curves), and
(d) PSR J1902+0556 at 430 MHz.
The best fit ISM models (PBFs) are shown as dashed curves
(panels $a$, $b$ and $d$), and the CLEANed pulses are shown
as thick solid curves.  The negative dip preceding the pulse
in (a) is an artifact that appears for several choices of PBF and
$\taud$.
}
\label{fig:examples}
\end{figure*}

Having demonstrated the feasibility of the technique using simulated
data, we now discuss its application to some real data. The data used 
in this paper are from
recent observations at Arecibo, where a large sample of high-DM
pulsars has been studied at multiple frequency bands \citep{bhat02}.
The sample includes several new discoveries from the Parkes multibeam
survey \citep{manchesteretal2001} that are visible at Arecibo.
Details of the observations and results will be presented 
in a forthcoming paper.  
Figure~\ref{fig:J1852} shows the 1475~MHz profile of PSR~J1852+0031 
(B1849+00), whose DM$= 680$~\dmu implies a distance D$= 8.4$~kpc
using the model of \citet{TC1993}.  
We attempted deconvolving these data with two widely different 
types of PBFs: (a) a sharp-edged PBF (eq.~\ref{eq:thin}) that characterizes
a thin slab model, (b) a rounded PBF (eq.~\ref{eq:uniform}) for a
uniform distribution of scattering material. As the figure shows,
the results are strikingly different; the final CLEAN pulse appears
approximately triple for an exponential PBF (with a
characteristic time $\taud$=225$\pm$14 ms), whereas it is a distinct
double for the case of a uniform medium ($\taud$=121$\pm$6
ms). At first glance, either of the models seem equally good.
However, given that two-component profiles are more commonly
observed at and near this radio frequency (which implies observations 
are more sensitive to a conal geometry than a core and cone case), 
we believe that a double profile is more likely.
The double profile obtained with a rounded PBF is constituted from
only 7 CCs, compared to 21 required to fit the data with
an exponential PBF. As alluded to before, this economy in the number
of CCs lends weight to the idea that the intrinsic profile is double.
Very similar results are obtained for the same pulsar at 1175~MHz
(Figure~\ref{fig:examples}a), which implies that although 
the folded profile at 1475~MHz is 
an average from only $\sim$120 pulse periods, the result is not
likely to be affected by inadequate profile stability.

More examples are shown in Figure~\ref{fig:examples}, where the data
span a wide range of signal to noise ratio and degree of scattering.
The CLEAN method yields satisfactory and robust results even for cases
where the scattering tail is not very prominent, as illustrated by the
scattering tail of PSR~J1902+0556, which extends only to a few percent
of the pulse period. The 1175 MHz profile of PSR~J1852+0031
illustrates the other extreme, where the scattering tail extends out
to nearly half the pulse period.  For the 430 MHz data on
PSR~J1901+0331, three different types of PBFs
(eqs.~\ref{eq:thin},~7~and~\ref{eq:uniform}) were used for deconvolution. As
evident from Figure~\ref{fig:examples}c, the observations are better
accounted for by a thin slab case with exponential PBF of
$\taud$=60$\pm$3 ms or by a uniform medium with a PBF of
characteristic time $\taud$=31$\pm$2 ms, rather than an intermediate
case such as a thick slab (eq.~\ref{eq:thick}).  The residual
asymmetry in the deconvolved pulse is most likely intrinsic to the
pulsar beam, though it is possible that the true PBF for this line of
sight differs from what we have considered.  Detailed interpretations
are deferred to a forthcoming paper where we will consider
multifrequency data and additional forms for the PBF.


\def\tnm{\tablenotemark}
\begin{deluxetable}{lcccccccc}
\tablecolumns{9}
\tablewidth{0pc}
\tablecaption{Pulse-broadening Times: Frequency
Extrapolation vs CLEAN \label{tab1}\tnm{a}}
\tablehead{
\colhead{} & \colhead{} & \colhead{} &
\colhead{} & \multicolumn{2}{c}{F-Extrapolation} &
\colhead{} & \multicolumn{2}{c}{CLEAN} \\
\cline{5-6} \cline{8-9}
\colhead{PSR} & \colhead{DM} & \colhead{Distance\tnm{b}} &
\colhead{Frequency} & \colhead{$\taud$} & \colhead{$\sigtd$} &
\colhead{} & \colhead{$\taud$} & \colhead{$\sigtd$} \\
\colhead{}      & \colhead{(\dmu)} & \colhead{(kpc)} &
\colhead{(MHz)} & \colhead{(ms)}   & \colhead{(ms)} &
\colhead{}      & \colhead{(ms)}   & \colhead{(ms)}
}
\startdata

J1852+0031    & 680   &  8.4 &1175  & 367 & 75 & & 487 & 73 \\
J1852+0031    & 680   &  8.4 &1475  & 190 & 30 & & 225\tnm{c} & 14 \\
J1853+0546    & 197.2 &  4.9 &1175  &  12 &  2 & &  13 & 2 \\
J1853+0546    & 197.2 &  4.9 &1475  &   6 &  1 & &   6 & 1 \\
J1855+0422    & 438.6 & 10.1 &1175  &  25 & 15 & &  24 & 4 \\
J1901+0331    & 401.2 &  7.8 & 430  &  57 &  3 & & 61 & 2 \\
J1902+0556    & 179.7 &  3.9 & 430  &  15 &  3 & &  14 & 1 \\

\enddata

\tablenotetext{a}{Fits, except where noted, used a pulse broadening
function (PBF) that is a one-sided exponential, appropriate for
a thin screen having a square-law structure function.}

\tablenotetext{b}{Derived from DM and the \citet{TC1993} model
for \nele in the Galaxy.}

\tablenotetext{c}{Data are equally consistent with a rounded
PBF (e.g. for a uniform medium) yielding $\taud=121\pm6$ ms
and a double pulse (Figure~\ref{fig:J1852})
instead of a merged triple pulse.}

\end{deluxetable}

\noindent
\subsection{Estimates of Pulse-broadening Time}
\label{s:est}
The CLEAN-based approach we outlined in this paper provides a general tool
that can be applied to a wide variety of measured pulse shapes and degrees
of scattering. For pulse shapes that 
show significant scattering tails, application of the frequency 
extrapolation approach as outlined in \S\ref{s:conv} can be expected to
yield satisfactory results. 
We have applied the CLEAN and frequency-extrapolation methods
to strongly scattered pulses, such as those in 
Figure~\ref{fig:examples}, and the results are 
tabulated in Table~\ref{tab1}. 
The data span a wide range of DM, signal 
to noise ratio and degree of scattering. In the frequency 
extrapolation method, the uncertainty in $\taud$ is estimated as 
the model parameter range where \chisq is unity above the minimum. 
For the CLEAN method, a similar estimate can be defined based on the
figure of merit parameter $f_r$, which is essentially an indicator 
of the average deviation of the residuals from the noise level 
(normalized to the noise variance). For instance, if a model
results in a value for $f_r$ that is unity above that for the 
best fit PBF, then the pulse is over-subtracted by \sigoff 
(on average). Note that the best fit model is characterized by 
a minimum in the combined figure of merit parameter $f_c$.


The CLEAN-based pulse-broadening measurements in Table~\ref{tab1} 
for PSRs J1852+0031 and J1853+0546 
indicate scaling laws $\taud \propto \nu^{-x}$ with
$x = 3.45\pm 0.7$ and $x = 3.4\pm 1.0$, respectively.
These scalings may be compared with those expected from a
Kolmogorov wavenumber spectrum with a negligibly small inner scale 
($x = 4.4$) or from a medium with a square-law
structure function ($x = 4$).  
The empirical intervals for $x$  include both cases, though the centroids
of the intervals are certainly biased below the expected values.
We defer a detailed analysis to a later paper, but we point out here
that the inferred scaling index for a given pulsar will depend on
the PBF used and that the PBF shape may also be a function of frequency.
Ideally, the data and the CLEAN procedure will indicate the best
PBF through investigation of the figures of merit discussed above. 
Both the PBF and the scaling index of $\taud$ are useful
information for understanding the nature and distribution of
scattering along the line of sight
\citep{CL2001,lohmer2001,boldyrev-gwinn2002}.  For multifrequency data
like these, the CLEAN-based approach we have outlined here can be
extended to do a global fit (by assuming the same form for the PBF,
and $\taud$ scaling with the frequency as a power law but with an
unknown index) to yield the pulse-broadening times and the index of
the power law, along with the intrinisic pulse shapes. A detailed
treatment of this approach is deferred to a forthcoming paper where we
report on multifrequency pulse-broadening measurements for a larger
sample of distant pulsars.

\section{Conclusions}
\label{s:conclusions}

We have developed and applied an algorithm for deconvolving
interstellar pulse broadening from radio pulses. The technique
realizes the two-fold objective of recovering the intrinsic pulse
shape and determining the pulse-broadening function that describes the
scattering process.  The method parallels the CLEAN algorithm used in
synthesis imaging applications, but differs from CLEAN in that the best
pulse broadening function is not known in advance. Additionally, we
perform a true deconvolution, where the resolution of the deconvolved
pulse is limited by the instrumental resolution rather than the pulse
broadening function. 

Application of the technique is demonstrated through use of simulated
data as well as data from Arecibo observations. Unlike the
conventional frequency extrapolation approach, the method makes no
prior assumptions about the intrinsic pulse shape, and can therefore
be applied to a wider variety of pulse shapes and degrees of
scattering.  

The derived intrinsic pulse shape depends on the form and parameters
of the PBF used for deconvolving, and further considerations have to
be employed to select the best fit model.  We employ figures of merit
that quantify positivity and asymmetry of the CLEANed pulse, and the
best fit PBF is taken to be the one that yields minimal residual
scattering and skewness of the intrinsic pulse. As in any other
nonlinear method, the uniqueness of the deconvolved result is not
always assured: however, further considerations (such as data at
different observing frequencies) can help resolve any such degeneracy.
Intrinsic pulse shapes obtained by application of this method are
likely to be closer to reality, and hence should prove to be useful
for the study of pulsar emission properties. A paper is in preparation
which applies our method to a large body of multi-frequency pulsar
observations. 

\acknowledgements 

We thank the referee, S. R. Spangler, for several thoughtful comments
that helped us to improve the paper.  This work was supported by NSF 
Grant AST~9819931 and by the National Astronomy and Ionosphere Center, 
which operates the Arecibo Observatory under cooperative agreement 
with the NSF.


\end{document}